# XML parser GUI using .NET Technology

## Seifedine Kadry[a,*], Jimbo Claver[a]

*American University of the Middle East, Kuwait*

**Abstract**

The purpose of this paper is to implement software that can save time, effort, and facilitate XML and XSL programming. The XML parser helps the programmer to determine whether the XML document is Well-formed or not, by specifying if any the positions of the errors.

## 1. Introduction

In the ever evolving world of programming, it becomes more challenging to make programming easier, faster and more efficient. XML (eXtensible Markup Language) is a meta-language in which other languages are created. XSL (eXtensible Style sheet Language) is used for formatting XML file [1, 2]. The goal of this paper is to design and implement a parser for XML, and a Graphical Interface to edit XSL files using .NET technology [3, 4].

This paper is organized as follows: In section 2, we provide an overview of XML. Then, section 3 describes XSL language. In section 4, the implementation of the XML parser is presented. Finally, the conclusion and future work.

## 2. Why XML?

XML is a markup language that is developed by the World Wide Web (WWW) Consortium to overcome the HTML (HyperText Markup Language) limitations, since XML is a flexible, scalable, and adaptable language. It allows itself to be used for communication between different systems, and for distributing information over the Web or between software applications without the constraints and eliminations of HTML. It is also characterized by: simplicity, openness, self-description, machine-readable context information, separation of content from presentation, multilingual documents and Unicode…..

---

\* Corresponding author. Tel.: 00965-66610985
*E-mail address:* skadry@gmail.com.

The XML programmer can create his own tags, elements and Markup Language that fits his needs. XML would adopt a very strict syntax which results in smaller, faster, and lighter browser.

XML is designed to describe data and documents. This data might be intended to be read by people or machines. It can be highly structured data such as data typically stored in databases or spreadsheets, or loosely structured data such as data stored in letters or manuals. It has gained acceptance in many industries as an integrating platform for applications and data sources. The programmer could start from large text tags, to paragraph, to subparagraph, to sentence tags, to URL tags and so forth. A Well-formed XML document must contain elements, attributes, text, and respect certain rules [5, 6].

Every XML document has a logical structure and a physical structure. A Logical structure defines the use of elements, attributes, data type, and other components employed in XML. The definition and declaration of the elements that make up the XML document are crucial to the success of the document and its viability with other documents. The physical structure provides the data that goes into the elements, such as text images, or other media, as allowed by the logical structure.

## 3. What is XSL?

XSL is used to apply XML styling to XML. This is accomplished using the eXtensible Style sheet Language Transformation (XSLT) in which XML is combined with CSS (cascade style sheet) to create a document that can be rendered on a Web-browser or other User Agent (UA).The process of styling requires an XML source document that contains the information to be displayed, and a style sheet to define and describe how the document should be presented in the XML document. There are Processing Instructions (PIs) that declare the XSL document to be used in CSS and its location or Uniform Resource Identifier (URI). XPath is used by XSLT as a mean to create a result tree from the source tree (XML document) and the instruction tree of the XSLT document. XPath discusses how a node can be specified, how the XML document tree may be traversed, and how the subsections of the document may be accessed. XSLT uses XPath to specify where information is stored in the XML file (source XPath is used again, but this time as notation from the XSLT file for styling information [7].

## 4. XML Parser

### *Design*

Our parser contains three buttons: textbox, rich textbox, and an array list. The textbox is used to specify the path of the XML file , the rich textbox is used to show the code of the XML file, and the array list is used to specify where the error is exactly (line and position) in an XML file that is not Well-formed.

The three buttons are:

    1-  Load file button: it is used to browse the XML file
    2-  Save as button: it is used to save the file after alteration
    3-  Validate button: it is used to check the XML file whether it is Well-formed or not

The path of the XML file should be specified or the file can be browsed. A Well-formed XML document helps to ensure that the information is structured in a way that is sensible for applications which use it. XML documents that are merely Well-formed can store any element inside of or adjacent to any other element.

The advantages of our XML parser are:

The development of XML documents stays separate from their programmatic modification. The content of the document remains legal XML that can be developed using standard interactive design tools

Parsing of the XML document is done at compile time rather than run time. This may offer a significant performance benefit for complex pages

The flow of control of the code remains separate from the page

It is helpful for generating dynamic XML pages on a server for display in a browser; it is a good choice for any application that requires manipulation of pre-parsed XML files.

In the following section, the implementation of a parser will be shown. This compiler was done using Csharp.net

Csharp.NET provides five namespaces to support XML classes. These namespaces are as follows:
- System.XML
- System.XML.Schema
- System.XML.Serialization
- System.XML.XPath
- System.XML.XSL

The System.XML namespace contains major XML classes. This namespace contains many classes to read and write XML documents. These classes are as follows:
- XMLReader
- XMLTextReader
- XMLValidatingReader
- XMLNodeReader
- XMLWriter
- XMLTextWriter

There are four reader and two writer classes, these reader and writer classes are used to read and write XML documents. The XMLReader class is an abstract base class which contains methods and properties to read a document. The Read method reads a node in the stream. Besides its reading functionality, this class also contains methods to navigate through a document nodes. The XMLWriter class can write data to XML documents. This class provides many writer methods to write XML document items. This class is base class for XMLTextWriter class. The XMLTextReader, XMLNodeReader and XMLValidatingReader classes are derived from the XMLReader class. Besides XMLReader methods and properties, these classes also contain members to read text, node, and schemas respectively. XMLTextReader class is used in the compiler to read an XML file by passing the file name as a parameter in constructor. XMLTextReader textReader = new XMLTextReader ("C:\\books.XML") . After creating an instance of the XMLText Reader, you call the Read method to start reading the document. After the read method is called, you can read all information and data stored in a document. XMLReader class has properties such as Name, BaseURI, Depth, LineNumber and so on [8, 9].

Implementation steps

1- Create an object of XML VALIDATOR: Create a class XMLValidator that uses many classes of XML in order to compile and handle the error.

```
String f = file.text;
XMLValidator Validator=new  XMLValidator(f);
```

f is the path of the file

2- Load the XML file: The file can be loaded using openFileDialog that can browse the file in order to select it.

```
openFileDialog1.ShowDialog();
```

```
string f;
f = openFileDialog1.FileName;
file.Text = f;
```
3-Read & show contents of the file: To read from the XML file, the ASCIIEncoding class should be used.
```
char[] charData = new Char[200];
ASCIIEncoding ae = new ASCIIEncoding();
byte[] b = new byte[200];
FileStream fs = new FileStream(@f, FileMode.OpenOrCreate);
fs.Read(b, 0, b.Length);
fs.Close();
Decoder d = Encoding.Default.GetDecoder();
d.GetChars(b, 0, b.Length, charData, 0);
string s = "";
for (int j = 0; j < charData.Length; j++)
s+= charData[j];
txtXML.Text = s;
```
4-Read the path: To read the path, the XMLTextReader class is used, and to read the content of the file the XMLValidatingReader class is used.
```
XMLTextReader txtreader = new XMLTextReader(fileName);
reader = new XMLValidatingReader(txtreader);
```
5-Handle the error: For handling the error, the ValidationEventHandler class is used; it determines the errors and put them in an array list.
```
reader.ValidationEventHandler += new ValidationEventHandler
(ValidationCallBack)
```
The compiler implemented above was tested with good (valid) and bad (not valid) XML documents. Both tests were successful; we can alter the document and save it. The examples shown below illustrate the tests done on the compiler. Note that you can either enter the path of the XML file or browse it then press validate.
Examples

*Example 1: 1st rule (opening and closing tag)*

XML documents must contain a unique opening and closing tag that contains the whole document, forming what is called a root element. In this example, the second column is not well formed because it lacks a root element as in the first column. (cf. Figure 1)

   <videocollection>...</videocollection>

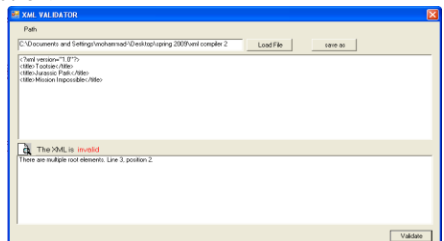

Figure 1

| Well-formed | Not Well-formed |
|---|---|
| <videocollection> | <title>Tootsie</title> |
| <title>Tootsie</title> | <title>Jurassic Park</title> |
| <title>Jurassic Park</title> | <title>Mission Impossible</title> |
| <title>Mission Impossible</title> | |

    </videocollection>>

*Example 2: 2nd rule (nested properly)*
Other than the root element, all other tags in an XML document must be nested properly, i.e. there must be an opening and a closing tag and the tags cannot overlap. In HTML, the tags would normally stand alone, such as  or <br> Tag are called "empty Tags". In XML, empty Tags look like this: e.g. : <BR/>. Whereas something like
</title... has no closing angle bracket; therefore the tag is not complete! And
</title)...has a wrong closing bracket, therefore the tag is not complete!
In this second example we show how the compiler deals with tags that are not properly nested. (cf. Figure 2)

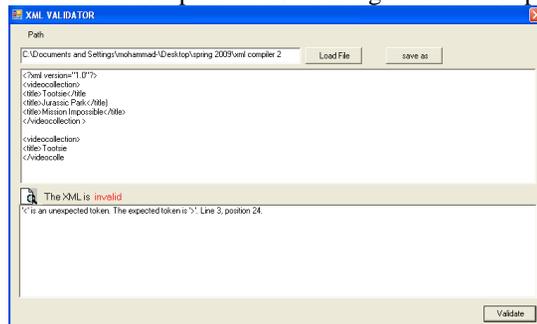

Figure 2

Well-formed

<videocollection>
<title>Tootsie</title>
<title>Jurassic Park</title>
<title>Mission Impossible</title>
</videocollection>
<videocollection>
<title>Tootsie</title>
</videocollection>

Not Well-formed

<videocollection>
<title>Tootsie</title
<title>Jurassic Park</title)
<title>Mission Impossible</title>
</videocollection >
<videocollection>
<title>Tootsie
</videocollection></title>

*Example 3: 3rd rule (case sensitive)*
Tags in XML are case sensitive, that means that <CREW>, <Crew> and <crew> are not the same. The XML processing instruction must be all lowercase, while keywords in DTDs must be all UPPERCASE, such as ELEMENT, ATTLIST, #REQUIRED, #IMPLIED, NMTOKEN, ID, etc. However, your own elements and attributes may be any case you choose, as long as they are consistent. Figure 3 shows the errors generated by the compiler when the case of the tags does not match.

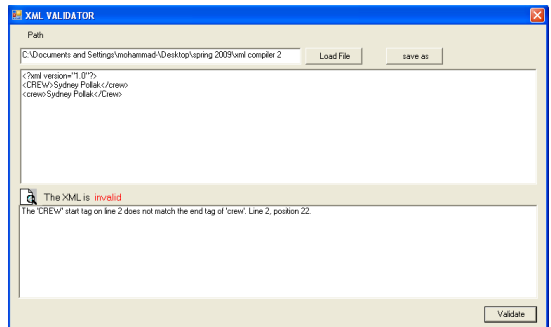

Figure 3

| Well-formed | Not Well-formed |
|---|---|
| <crew>Sydney Pollak</crew> | <CREW>Sydney Pollak</crew> |
| | <crew>Sydney Pollak</Crew> |

*Example 4:  4th rule (quoted attributes)*

As opposed to HTML, the attribute values in XML must always be quoted. Figure 4 shows that the compiler returns errors when the attribute is not quoted.

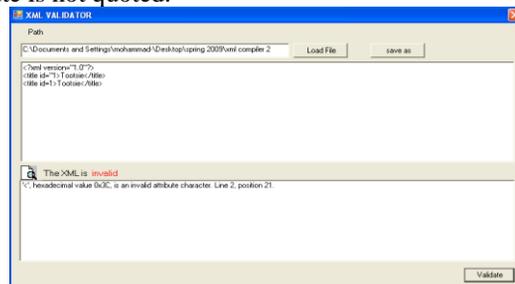

Figure 4

| Well-formed | Not Well-formed |
|---|---|
| | <title id="1">Tootsie</title> |
| <title id="1">Tootsie</title> | <title id=1>Tootsie</title> |

*Example 5: 5th rule (declaration)*

Another rule in XML is to begin the XML document with a declaration. Errors during compilation would occur if the declaration is missed. This is shown in the following figure. (cf. Figure 5)

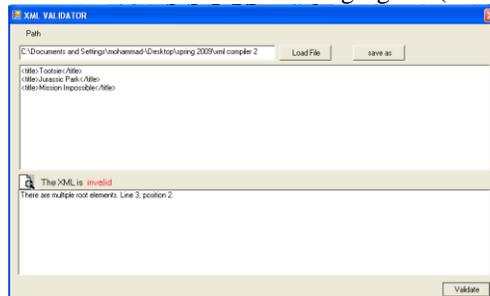

Figure 5

Well-formed                                         Not Well-formed
<?XMLversion="1.0"?>                               <title>tootsize</title>
<title>tootsize</title>                             <title>Jurassic Park</title>
<title>Jurassic Park</title>                        <title>Mission Impossible</title>
<title>Mission Impossible</title>

## Conclusion

In this paper, we introduced our XML parser that was implemented using XML classes. The main goal of this parser was to check XML documents for errors in a very easy and fast way, in order to help the programmer to determine whether the XML document is Well-formed or not. As stated earlier, this parser was designed using XML classes. These classes use XML functions which are very efficient in compiling and handling errors in XML documents.